\documentclass[sigconf]{acmart}
\AtBeginDocument{%
  }

\setcopyright{acmlicensed}

\copyrightyear{2025} 
\acmYear{2025} 
\setcopyright{acmlicensed}\acmConference[MM '25]{Proceedings of the 33rd ACM International Conference on Multimedia}{October 27--31, 2025}{Dublin, Ireland}
\acmBooktitle{Proceedings of the 33rd ACM International Conference on Multimedia (MM '25), October 27--31, 2025, Dublin, Ireland}
\acmDOI{10.1145/3746027.3754734}
\acmISBN{979-8-4007-2035-2/2025/10}

\usepackage{balance} 
\begin{document}

\title{FlowDubber: Movie Dubbing with LLM-based Semantic-aware Learning and Flow Matching based Voice Enhancing}

\newcommand{\dubber}{FlowDubber}

\author{Gaoxiang Cong}
\orcid{0009-0002-1239-8119}
\affiliation{%
  \institution{Institute of Computing Technology, Chinese Academy of Sciences \& University of Chinese Academy of Sciences}
  \city{Beijing}
  \country{China}}
\email{gaoxiang.cong@vipl.ict.ac.cn}

\author{Liang Li}
\orcid{0000-0002-1943-8219}
\authornote{Corresponding author.}
\affiliation{%
  \institution{Institute of Computing Technology, Chinese Academy of Sciences}
  \city{Beijing}
  \country{China}}
\email{liang.li@ict.ac.cn}

\author{Jiadong Pan}
\orcid{0000-0002-4436-8830}
\authornote{Equal contribution.}
\affiliation{%
  \institution{Institute of Computing Technology, Chinese Academy of Sciences \& University of Chinese Academy of Sciences}
  \city{Beijing}
  \country{China}}
\email{panjiadong18@mails.ucas.ac.cn}

\author{Zhedong Zhang}
\orcid{0009-0003-1129-8914}
\affiliation{%
 \institution{Hangzhou Dianzi University}
 \city{Hangzhou}
 \country{China}}
\email{zhedong_zhang@hdu.edu.cn}

\author{Amin Beheshti}
\orcid{0000-0002-5988-5494}
\affiliation{%
  \institution{Macquarie University}
  \city{Sydney}
  \country{Australia}}
\email{amin.beheshti@mq.edu.au}

\author{Anton van den Hengel}
\orcid{0000-0003-3027-8364}
\affiliation{%
  \institution{University of Adelaide}
  \city{Adelaide}
  \country{Australia}}
\email{anton.vandenhengel@adelaide.edu.au}

\author{Yuankai Qi}
\orcid{0000-0003-4312-5682}
\affiliation{%
  \institution{Macquarie University}
  \city{Sydney}
  \country{Australia}}
\email{qykshr@gmail.com}

\author{Qingming Huang}
\orcid{0009-0006-8793-6953}
\affiliation{%
  \institution{University of Chinese Academy of Sciences}
  \city{Beijing}
  \country{China}}
\email{qmhuang@ucas.ac.cn}

\def\eg{\emph{e.g.}} 
\def\Eg{\emph{E.g.}}
\def\vs{\emph{v.s.}} 
\def\ie{\emph{i.e.}} 
\def\Ie{\emph{I.e.}}
\def\etc{\emph{etc.}} 
\def\wrt{\emph{w.r.t.}} 
\def\etal{\emph{et al.}}

\renewcommand{\shortauthors}{Gaoxiang Cong et al.}

\begin{abstract}
 
Movie Dubbing aims to convert scripts into speeches that align with the given movie clip in both temporal and emotional aspects while preserving the vocal timbre of a given brief reference audio. 
Existing methods focus primarily on reducing the word error rate while ignoring the importance of lip-sync and acoustic quality. 
To address these issues, we propose a novel dubbing architecture based on Large Language Model (LLM) and Conditional Flow Matching (CFM), named FlowDubber, which achieves high-quality audio-visual sync and pronunciation by incorporating a large speech language model with dual contrastive alignment while improving acoustic quality via Flow-based Voice Enhancing (FVE). 
First, we introduce Qwen2.5 as the backbone of large speech language model to learn the in-context sequence from movie scripts and reference audio. 
Second, the proposed semantic-aware learning focuses on capturing LLM semantic knowledge at the phoneme level, which facilitates mutual alignment with lip movement from silent video via Dual Contrastive Alignment (DCA). 
Third, the FVE introduces an LLM-based acoustics flow matching guidance to strengthen clarity by decoupling Classifier-Free Guidance (CFG) enhancement. 
Extensive experiments demonstrate that our method outperforms several state-of-the-art methods on two primary benchmarks. 
The demos are available at   {\href{https://galaxycong.github.io/LLM-Flow-Dubber/}{\textcolor{blue}{https://galaxycong.github.io/LLM-Flow-Dubber/}}}.

\end{abstract}

\ccsdesc[500]{Computing methodologies~Phonology / morphology}
\ccsdesc[300]{Computing methodologies~Computer vision}

\keywords{Movie Dubbing, Visual Voice Cloning, Flow Matching}


\maketitle


\section{Introduction}

\begin{figure}
    \centering
    \includegraphics[width=1\linewidth]{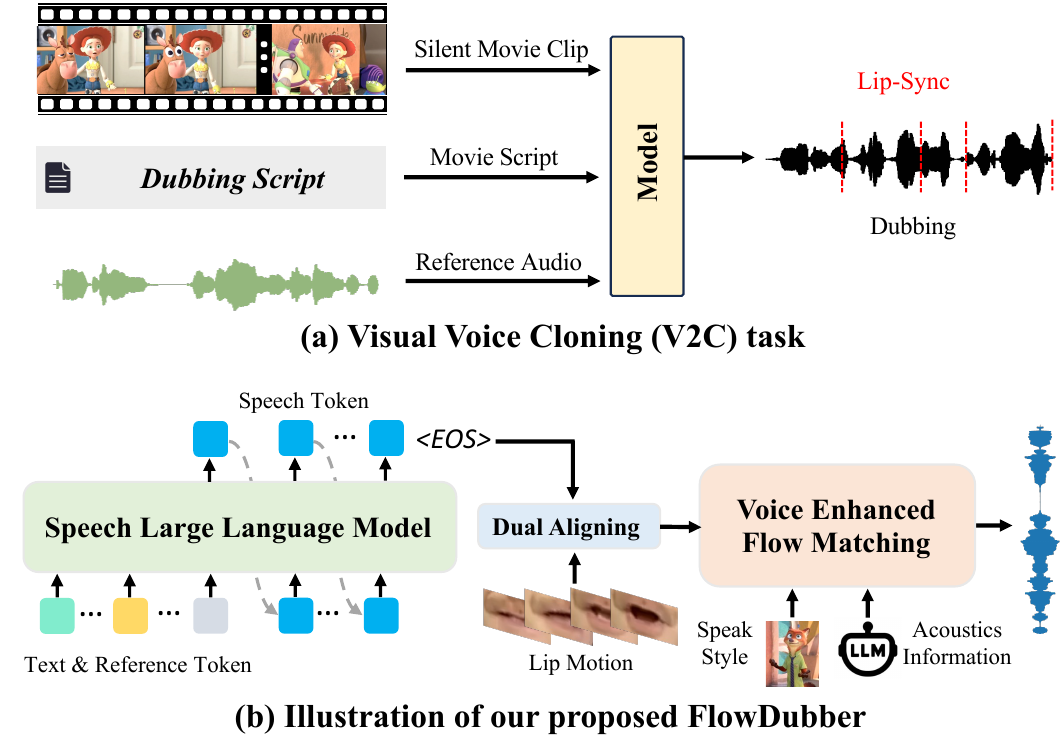}
    \caption{(a) V2C task. (b) Unlike dubbing with duration predictor from TTS, FlowDubber incorporates a large language model (LLM) and voice-enhanced flow matching to generate high-quality dubbing while ensuring lip-sync. }
    \label{fig:intro}
    \vspace{-0.4cm}
\end{figure}

Movie Dubbing, also known as Visual Voice Cloning (V2C)~\cite{chen2022v2c}, aims to generate a vivid speech from scripts using a specified timbre conditioned by a single short reference audio while ensuring strict audio-visual synchronization with lip movement from silent video, as shown in Figure~\ref{fig:intro}(a). 
It attracts great attention in the multimedia community and promises significant potential in real-world applications such as film post-production and personal speech AIGC.

Previous dubbing works~\cite{zhang2024speaker, zhang2025produbber, cong2024styledubber, cong2023learning, chen2022v2c} achieve significant progress in improving pronunciation and are dedicated to reducing the word error rate (WER) of generated speech. %
They can be mainly divided into two groups. 
Since the dubbing resources are limited in scale (copyright issues) and are always accompanied by background sounds or environmental noise, one class of methods~\cite{zhang2024speaker, zhang2025produbber} focuses primarily on leveraging external knowledge to improve pronunciation clarity by pre-training on clear, large-scale text-to-speech corpus~\cite{HeigaLibriTTS}. %
For example, Speaker2Dubber~\cite{zhang2024speaker} proposes a two-stage dubbing architecture, which allows the model to first learn pronunciation via multi-task speaker pre-training on Libri-TTS 100 dataset and then optimize duration in stage two. 
Then, by pre-training on larger TTS corpus Libri-TTS 460 dataset, ProDubber~\cite{zhang2025produbber} proposes another novel two-stage dubbing method based on the Style-TTS2 model~\cite{YinghaoStyleTTS2}, including prosody-enhanced pre-training and acoustic-disentangled prosody adapting.  
However, these pre-training methods rely too much on the TTS architecture~\cite{ren2020fastspeech, YinghaoStyleTTS2} and mainly adopt a Duration Predictor (DP)~\cite{cong2024styledubber} to produce rough duration without considering intrinsic relevance with lip motion, resulting in poor audio-visual sync. %

The other family of methods~\cite{chen2022v2c, cong2023learning, cong2024styledubber} do not care about pre-training, but try to decline WER by associating other related modality information that helps with pronunciation. 
For example, Styledubber~\cite{cong2024styledubber} proposes a multi-modal style adaptor to learn pronunciation style from the reference audio and generate intermediate representations informed by the facial emotion presented in the video. 
However, due to the introduction of time stretching, StyleDubber~\cite{cong2024styledubber} can only keep the global time alignment (\ie, the total length of the synthesized dubbing is consistent with the target), which is still unsatisfactory in fine-grained matching with lip motion, bringing a bad audio-visual experience.

Except for the alignment issues mentioned above, the existing dubbing methods suffer from acoustic quality degradation, even in the advanced two-stage dubbing pre-training methods. 
For example, Speaker2Dubber~\cite{zhang2024speaker} freezes the text encoder in the second stage, which helps to maintain pronunciation.  
However, its use of a traditional FastSpeech2-based~\cite{ren2020fastspeech} transformer fails to handle the complex and diverse spectrum changes, leading to subpar acoustic quality. 
In addition, the acoustic quality measurement predictor UTMOS~\cite{utmos} reveals that the acoustic quality of current dubbing methods still requires improvement.

Recent advances in speech tokenization~\cite{OordNeural, du2024cosyvoice2,
FiniteFabianMentzer, ji2024wavtokenizer} have revolutionized TTS synthesis by bridging the fundamental gap between continuous speech signals and token-based large language models (LLM).  
Due to LLM demonstrating excellent capability in sequential modeling and contextual understanding, these LLM-based speech synthesis models achieve human-level expressive and naturalness~\cite{du2024cosyvoice2, llasa, wang2025spark, guo2024fireredtts}. 
However, they are struggling to deal with the dubbing task. 
Although some speed-controllable LLM speech models have been proposed, they still lack visual understanding capabilities, and the synthesized speech struggles to align with the lip motion changing in video.

To address these issues, we propose an LLM-based flow matching architecture for dubbing, named FlowDubber (as shown in Figure~\ref{fig:intro} (b)), which incorporates a large speech language model and dual contrastive alignment to ensure audio-visual sync and pronunciation,  while achieving better acoustic quality via voice-enhanced flow matching than the state-of-the-art method. 
Specifically, we first introduce an LLM-based Semantic-aware Learning (LLM-SL), which leverages pre-trained LLM Qwen2.5-0.5B to model the in-context sequence from movie scripts (text) and reference audio (reference token including ref. semantic and ref. global token). 
Then, the proposed semantic-aware phoneme learning captures the connection between phoneme-level pronunciations and LLM-derived semantics, making them well-suited for integration into the Dual Contrastive Aligning (DCA) module.
Next, the DCA is designed to perform mutual alignment between lip movement and phoneme sequence to ensure lip-sync. 
Finally, we propose a novel Flow-based Voice Enhancing (FVE) module, which improves the acoustic quality from two sub-components: LLM-based acoustics flow matching guidance and style flow matching prediction.  
The key part is LLM-based acoustics flow matching guidance, which focuses on improving clarity during recovering noise to mel-spectrograms by decoupling Classifier-Free Guidance (CFG) enhancement.

The main contributions of the paper are as follows: 
\begin{itemize}
    \item  We propose a powerful dubbing architecture \dubber, which incorporates LLM for semantic learning and flow matching for acoustic modeling to enable high-quality dubbing, including lip-sync, acoustic clarity, speaker similarity. %
    \item We devise an LLM-based Semantic-aware Learning (LLM-SL) to absorb token-level semantic knowledge, which is convenient to achieve precisely lip-sync for dubbing by associating proposed dual contrastive aligning. %
    \item We design a Flow-based Voice Enhancing mechanism to enhance the semantic information from LLM, refining the flow-matching generation process for high speech clarity. 
    \item Extensive experimental results demonstrate the proposed \dubber~ performs favorably against state-of-the-art models on two popular dubbing benchmark datasets. %
\end{itemize}

\section{Related Work}

\subsection{Visual Voice Cloning} 
With the rapid development of deep learning~\cite{YunbinSMART, XuejingEntity, BeichenInductive, ye2022unsupervised, ZhangTaoDeepGuided, LiangPAMI}, V2C~\cite{chen2022v2c} has attracted great interest in the multimedia community~\cite{
choi2025v2sflow, choi2024av2av, transcpTang, vpp-llava, ZhedongGenerating, cui2024stochastic, JiaxinICASSP, ye2025emotional}. 
It requires determining how a text should be spoken, in sync with the lip movements in silent video and in the vocal style of reference audio~\cite{kim2025faces, sung2025voicecraft, lee2023imaginary, ye2025shushing, zhang2025deepaudio, zheng2025deepdubber, li2025fccondubber}. %
Some V2C works focus primarily on improving the pronunciation clarity~\cite{cong2023learning, cong2024styledubber, zhao2024mcdubber}. %
For example, SOTA dubbing method ProDubber~\cite{zhang2025produbber} and Speak2Dub~\cite{zhang2024speaker} propose a two-stage framework to learn clear pronunciation by pre-training from large-scale TTS corpus~\cite{HeigaLibriTTS}.  %
However, they over-rely on the TTS architecture and use an inaccurate duration predictor~\cite{zhang2025produbber} to estimate the lip speaking time, without considering the intrinsic audio-visual alignment. 
Besides, StyleDubber~\cite{cong2024styledubber} uses time stretching in the duration predictor. 
Although the overall length of the dubbing can be consistent, it does not fundamentally capture fine-grained lip-sync with the video. 
In this work, we propose \dubber, a novel dubbing architecture that combines LLM-based semantic-aware learning with dual contrastive alignment to achieve high-quality lip synchronization, and flow-matching enhancing mechanism is designed to achieve better acoustic quality than existing dubbing methods.

\begin{figure*}
    \centering
    \includegraphics[width=1\linewidth]{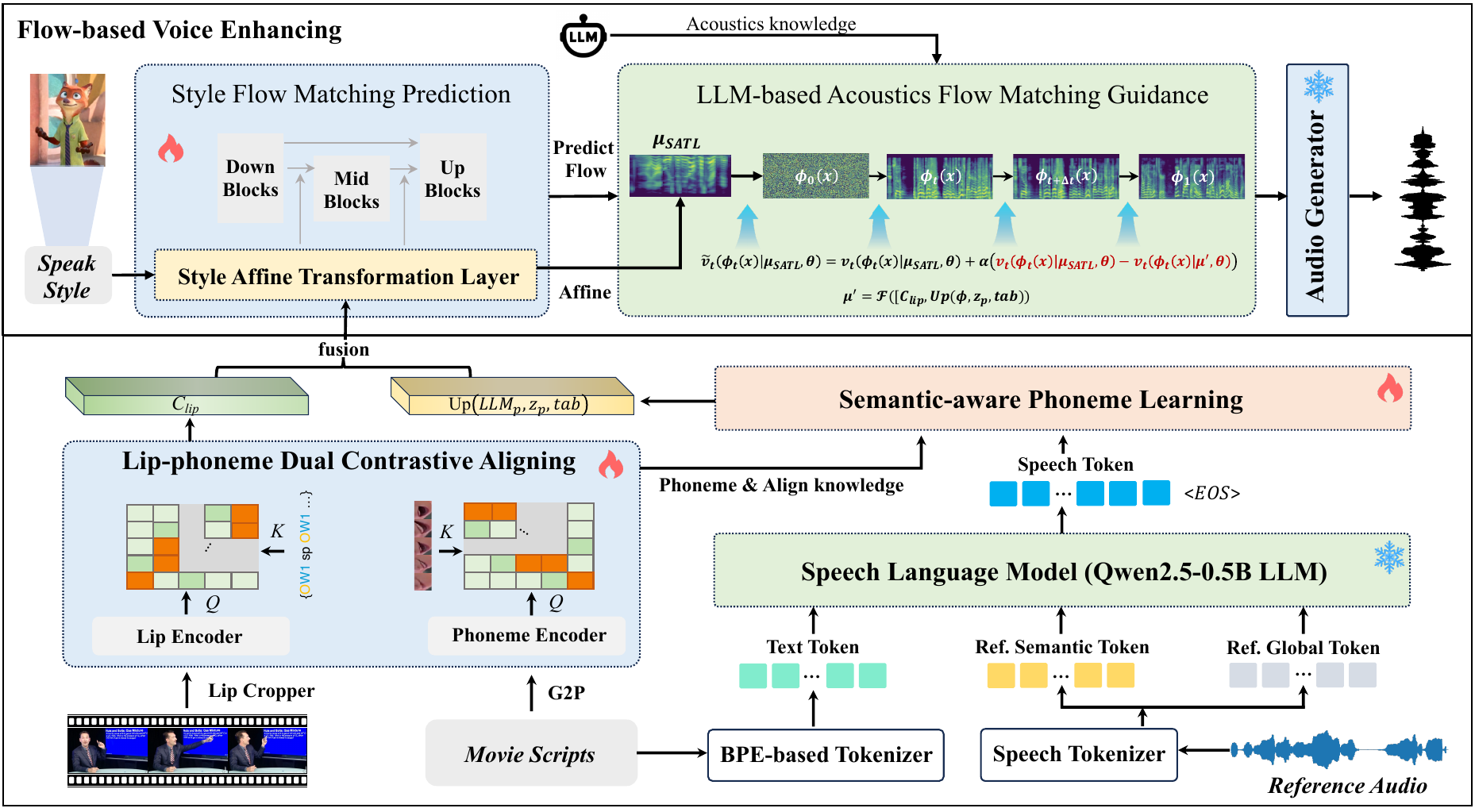}
    \caption{Overall framework of \dubber. It consists of LLM-based Semantic-aware Learning (LLM-SL),  Dual Contrastive Aligning (DCA), and Flow-based Voice Enhancing (FVE). 
    Specifically, the LLM-SL includes Qwen2.5-0.5B speech language model and semantic-aware phoneme learning to keep pronunciation while ensuring lip-sync by DCA.  
    The FVE is equipped with LLM-based Acoustics Flow Matching Guidance and Style Flow Matching Prediction to improve clarity and similarity. 
    } 
    \label{fig:archi}
\end{figure*}

\subsection{Large Language Model and Speech Codec} 
The remarkable success of Large Language Models (LLMs)~\cite{gpt, deepseekr1, qwen2} and the autoregressive (AR) model brings significant advancements in the field of speech synthesis.
VALL-E~\cite{vall-e} first converts speech into neural codec tokens and treats the speech synthesis as a next-token prediction task.
Subsequently, extensive research focuses on speech codecs and LLM-based speech generators to improve the synthesis performance.
For example, DAC~\cite{DAC} adopts the residual vector quantization and the multi-scale STFT discriminators to obtain higher-quality discrete speech tokens.
Wavtokenizer~\cite{wavtokenizer} and X-codec~\cite{xcodec} further improved the efficiency of codec and addressed the semantic shortcomings of previous codes.
Besides, LLM-based speech synthesis systems combine the AR model with other components~\cite{seedtts, valle2} or rely on continuous acoustic features~\cite{melle, kalle} to achieve better performance.
Recently, Llasa~\cite{llasa} investigated the effects of training-time inference-time scaling in LLM-based speech synthesis.
However, they still lack visual understanding capability, and the generated speech struggles to align with the lip movement. 
In this paper, we propose an effective dubbing model that can achieve high-quality audio-visual alignment and inherit the acoustic knowledge from LLM via Semantic-aware Phoneme Learning and  LLM-based Acoustics Flow Matching Guidance.

\subsection{Speech Synthesis and Flow Matching} 
Flow Matching~\cite{lipman2022flow} is a simulation-free approach to training continuous normalizing flow~\cite{chen2018neural} models, capable of modeling arbitrary probability paths and capturing the trajectories represented by diffusion processes~\cite{song2021maximum}. 
Due to the high quality and faster speed, flow matching has attracted significant attention in speech generation~\cite{YiweiVoiceFlow, SungwonFlow, VoiceboxMatthew, zhang-etal-2025}. 
Matcha-TTS~\cite{machaTTS} adopts the optimal transport conditional flow matching in single speaker TTS synthesis, and Stable-VC~\cite{yao2024stablevc} adopts it in voice conversion field to improve fidelity.  
F5-TTS~\cite{chen2024f5} is another powerful TTS model to reconstruct high-quality mel-spectrograms by flow matching. 
Then, CosyVoice 2.0~\cite{du2024cosyvoice1, du2024cosyvoice2} has further proven its superior performance by combining flow matching with LLM. 
However, these methods are not suited to the V2C dubbing task due to their inability to perceive proper pauses in step with lip motion. 
Recently, EmoDub~\cite{cong2024emodubber} introduces classifier guidance in flow matching to control emotions via input labels and intensity. 
In contrast, after integrating semantic-aware phoneme learning and lip-motion aligning, we focus on refining the flow-matching generation process to ensure clarity by introducing semantic knowledge from LLM via classifier-free guidance.

\section{Methods}

\subsection{Overview} 

The target of the overall movie dubbing task is:
\begin{equation}
    \hat{{Y}} = \mathrm{\dubber}(W_r, T_c, V_{s}),  
\end{equation} 
where the $V_{s}$ represents the given silent video clip, $W_r$ is a reference waveform used for voice cloning, and $T_c$ is current piece of text to convey speech content. 
The goal of \dubber~ is to generate a piece of high-quality speech $\hat{Y}$ that guarantees precise lip-sync with silent video, high speaker similarity, and clear pronunciation. 
The main architecture of the proposed model is shown in Figure~\ref{fig:archi}.
Specifically, we introduce pre-trained textual LLM Qwen2.5-0.5B as the backbone of the speech language model to learn the in-context sequence from movie scripts and reference audio by discretizing them. 
Then, the semantic knowledge of speech tokens is adapted to the phoneme level by semantic-aware phoneme learning.  
Next, the proposed Dual Contrastive Aligning (DCA) ensures the mutual alignment between lip-motion and phoneme-level information from LLM. 
Finally, Flow-based Voice Enhancing (FVE) aims to maintain the speaker's similarity and improve the clarity by an LLM-based Acoustics Flow Matching Guidance. 
We detail each module below.

\subsection{LLM-based Semantic-aware  Learning}\label{methods:1x} 
Different from the previous dubbing works~\cite{cong2023learning, zhao2024mcdubber}, we introduce LLM-based semantic-aware learning to capture the phoneme-level pronunciation via the powerful in-context learning capabilities of LLM (Qwen2.5-0.5B) between text tokens in movie script and semantic and global tokens in reference audio.

\noindent\textbf{Speech Tokenization}. 
This module aims to transform the speech signal of reference audio $R_a$ into a sequence of semantic tokens ${h}_q$, following Spark-TTS~\cite{wang2025spark}. 
It first utilizes a pre-trained self-supervised learning (SSL) model, wav2vec 2.0~\cite{wav2vecAlexei}, to translate speech signals into a semantic embedding sequence. 
Then, the semantic encoder $S_{encoder}(\cdot)$, constructed with 12 ConvNeXt~\cite{ZhuangConvNet} blocks and 2 downsampling blocks, is employed to process and down-sample the sequence further into an encoding sequence $h$: 
\begin{equation}
{H}_q=\text{VQ}({h}), {h}=S_{encoder}(\text{wav2vec2.0}(R_a)), 
\end{equation} 
where the output ${H}_q$ represents semantic tokens. 
$\text{VQ}(\cdot)$ adopts a factorized code structure with a codebook size of 8192 and 8 codebook dimensions. 
${G}_q$ denotes the global tokens by Finite Scalar Quantization (FSQ), following Spark-TTS. ~\cite{wang2025spark}.

\noindent\textbf{Speech Language Model}. 
Inspired by LLM successes, we employ the pre-trained Qwen2.5-0.5B~\cite{wang2025spark} as the backbone of the speech language model. 
Specifically, we formulate the GPT~\cite{radford2019language} architecture as the next-token prediction paradigm, which adopts a decoder-only autoregressive transformer architecture: 
\begin{equation}
    P(o_{1:N_o})=\prod_{i=1}^{N_o}P(o_i|T_q, H_q, G_q, o_1,\cdots,o_{i-1}),
\label{eq:ar}
\end{equation} 
where $o_i$ is the i-th generated speech token, and $N_o$ is the length of generated speech tokens. 
The $T_q$ represents text tokens by converting raw text $T_c$ using a byte pair encoding (BPE)-based tokenizer. 
$H_q$ are semantic tokens and $G_q$ are global tokens from reference audio.  
By inputting the concatenation of $T_q$, $G_q$, $H_q$ and previous special tokens $(o_{1}, ..., o_{i-1})$, model can autoregressively generate current speech tokens $o_i$ with in-context semantic knowledge.

\noindent\textbf{Phoneme Level Semantic-aware Module}. 
Compared with zero-shot TTS, movie dubbing must be strictly matched with lip movements from silent video to achieve audio-visual synchronization. 
The proposed phoneme-level semantic-aware module aims to capture the semantic knowledge from the speech language model at the phoneme level, which helps preserve pronunciation and enables fine-grained alignment between phoneme unit and lip motion sequence. 
Specifically, the phoneme-level semantic-aware module consists of cross-modal transformers $\hat{Z}^{[i]}_{S \rightarrow P}$ to calculate the relevance between textual phoneme embedding and LLM speech knowledge, which can be formulated as: 
\begin{equation}
    \begin{split}
        \hat{Z}^{[i]}_{S \rightarrow P} = & \; \text{LLM}^{[i],\text{mul}}_{S \rightarrow P} (\text{LN}(Z^{[i-1]}_{S \rightarrow P}), \text{LN}(Z^{[0]}_S))
        + \text{LN}(Z^{[i-1]}_{S \rightarrow P}), \\
        Z^{[i]}_{S \rightarrow P} = & \; f_{\theta^{[i]}_{S \rightarrow P}} (\text{LN}(\hat{Z}^{[i]}_{S \rightarrow P}) + \text{LN}(\hat{Z}^{[i]}_{S \rightarrow P}),
    \end{split}
\end{equation} 
where ${\text{LN}(\cdot)}$ denotes the layer normalization in cross modal transformer, $i = \{1,...,D\}$ denotes the number of feed-forwardly layers, and $f_{\theta}$ is a position-wise feed-forward sublayer parametrized by $\theta$. 
${\text{LLM}^{[i],\text{mul}}_{A \rightarrow L}}(\cdot)$ is a multi-head attention as follows: 
\begin{equation}
    \text{LLM}^{[i],\text{mul}}_{S \rightarrow P} =  \mathrm{softmax}(\frac{{E_{pho}(\text{G2P}(T_c))} {S_{llm}}^\top}{\sqrt{d_{m}}}){S_{llm}},
\end{equation} 
where $\text{G2P}(\cdot)$ denotes the grapheme-to-phoneme to convert raw text $T_c$ to a phoneme sequence, then the phoneme encoder $E_{pho}(\cdot)$ is used to obtain textual phoneme embedding.  
The $S_{llm}$ indicates the mapping speech feature from LLM token sequence $o_{1:N_o}$ by codec decoder~\cite{wang2025spark}. 
In this case, the $S_{llm}$ is used as key and value, and the textual phoneme embedding is used as query. 
Finally, we denote the last layer output of cross modal transformer as ${LLM}_p \in \mathbb{R}^{{l_p}\times{d_{m}}}$, which represents the phoneme-level semantic feature from LLM.  
The $l_p$ denotes the length of phoneme sequences and $d_{m}$ is the embedding size.

\subsection{Dual Contrastive Aligning for Dubbing}\label{methods:2x} 
This module is designed to achieve mutual alignment between lip movement sequence and phoneme sequence by introducing a dual contrastive learning after LLM-based Semantic-aware Learning. 

\noindent\textbf{Lip-motion Feature Extractor}. 
To ensure fairness for measuring alignment, we first use the same extractor~\cite{cong2023learning} to obtain lip motion features from silent videos $V_{s}$: 
\begin{equation}
    \begin{aligned}
        z_m=  \mathrm{LipEncoder}(\mathrm{LipCrop}(V_{s})), 
     \end{aligned}
\end{equation}
where $z_m\in\mathbb{R}^{{L_v}\times{d_{m}}}$ denotes the output lip motion embedding,  $L_v$ indicates the length of lip sequence, and $d_{m}$ is embedding size. 
The $\mathrm{LipCrop}(\cdot)$ uses the face landmarks tool to crop mouth area, and $\mathrm{LipEncoder}(\cdot)$ represents the lip encoder.

\noindent\textbf{Dual Contrastive Learning}. 
We focus on learning the intrinsic correlation between phoneme-level pronunciation and lip movement to achieve reasonable alignment for movie dubbing. 
Following the contrastive learning manner, we introduce the InfoNCE loss~\cite{YuandongUnderstanding} to encourage the model to distinguish correct lip-phoneme pairs. 
Specifically, we first treat the lip motion features $z_m$ as queries and the phoneme embeddings $z_p$ as keys. 
To establish positive pairs, we align each lip motion frame with its corresponding phoneme based on ground-truth timing annotations by Montreal Forced Aligner~\cite{mcauliffe2017montreal} (MFA) and Frames Per Second (FPS). 
This ensures that each $z_m^i$  should be maximally similar to its temporally aligned $z_p^j$, while being distinct from other phonemes: 
\begin{equation}
\mathcal{L}_{mp}=-\sum_i\log\frac{\sum_{j\in+}\exp(z_m^i\cdot z_p^j/\tau)}{\sum_j\exp(z_m^i\cdot z_p^j/\tau)},
\end{equation}
where $i\in[0, L_v-1]$ represents the $i$-th frame of the lip sequence and $j\in[0, L_t-1]$ represents the $j$-th textual phoneme from whole sequence. 
The $^+$ means positive sample pairs, which are calculated in advance based on the ground-truth information during training~\cite{cong2024emodubber}.  
Conversely, we introduce a second contrastive loss by reversing the roles: treating phoneme features $z_p$ as queries and lip motion embeddings $z_l$ as keys. 
In this case, each phoneme seeks to retrieve its temporally aligned lip feature while suppressing mismatched lip frames: 
\begin{equation}
\mathcal{L}_{pm}=-\sum_j\log\frac{\sum_{i\in+}\exp(z_p^j\cdot z_m^i/\tau)}{\sum_i\exp(z_p^j\cdot z_m^i/\tau)},
\end{equation}
unlike DLCL in Emodub~\cite{cong2024emodubber}, which focuses on aligning prosody sequences (obtained by prosody adaptor) to the other (lip), our method emphasizes aligning manner between original phoneme sequences and lip to reduce the impact of prosody changes. 
Besides, different from the single-direction aligning in~\cite{LiangPAMI}, our method focuses on a mutual aligning manner and does not rely on an extra duration predictor that learn coarse-grained time relevance by additional MSE loss. 
Finally, we use the average of mutual aligning results as dual contrastive loss: 
\begin{equation}
\mathcal{L}_{dua}= \frac{1}{2}\mathcal{L}_{mp} + \frac{1}{2}\mathcal{L}_{pm}. 
\end{equation}

\noindent\textbf{Aligning Phoneme Level Feature}. 
The similarity matrix between phoneme embedding and lip motion embedding ${Sim}(z_m, z_p)$ is constrained by dual contrastive learning, then ${Sim}(z_m, z_p)$ further guides the hybrid generation of aligned sequences, including:  
(1) lip-related aligning sequences $C_{lip}$.  
(2) phoneme related aligning sequences. 
Specifically, $C_{lip}$ is obtained by multi-head attention module in ~\cite{cong2023learning}, in which the $z_p$ serves as key and value, and the $z_m$ is the query. 
Unlike~\cite{cong2023learning}, the learnable ${Sim}(z_m, z_p)$ is used as multi-head attention weight matrix to provide correct relevance. 
Next, by monotonic alignment search (MAS)~\cite{JaehyeonGlow}, the ${Sim}(z_m, z_p)\in\mathbb{R}^{{L_v}\times{L_t}}$ is flat to mapping table $tab\in\mathbb{R}^{{L_t}\times{1}}$, which records the number of video frames corresponding to each phoneme unit. 
Finally, the $tab$, $LLM_p$, $z_p$, and $C_{lip}$ are associated to mel-spectrograms level prior conditions $\mu$: 
\begin{equation}
\mu =\mathcal{F}([C_{lip}, \mathrm{Up}(LLM_p, z_p, tab)), 
\label{eq:mu}
\end{equation} 
where $\mathrm{Up}(\cdot)$ is used to expand $LLM_p$ and $z_p$ to video level according to mapping $tab$. 
The $\mathcal{F}(\cdot)$ indicates the fusion module, which consists of 2D upsampling convolutional layers and transformer-based mel-decoder. 
The output $\mu \in\mathbb{R}^{{L_m}\times{d_m}}$, where ${l_m}$ and $d_m$ represent the length and embedding size of the target mel-spectrogram.

\subsection{Flow-based Voice Enhancing}\label{methods:3x}

In this section, we introduce flow-based voice enhancing, including Style Flow Matching Prediction to inject speaker style into flow matching and LLM-based Acoustics Flow Matching Guidance to improve the clarity of generated speech via decoupled Classifier-Free Guidance (CFG) enhancement.

\begin{table*}[!t]
  \centering
    \caption{
    Compared with related Dubbing methods on Chem benchmark. 
    For the Dub 1.0 setting, we use the ground truth audio as reference audio, for the Dub 2.0 setting, we use the non-ground truth audio from the same speaker within the dataset as the reference audio, which is more aligned with practical usage in dubbing. 
    $\uparrow(\downarrow)$ means that higher (lower) value is better. 
    }
    \vspace{-6pt}
  \resizebox{1.0\linewidth}{!}
  {
    \begin{tabular}{c|ccccc|ccccc}
    \hline
    Setting & \multicolumn{5}{c|}{Dubbing Setting 1.0} & \multicolumn{5}{c}{Dubbing Setting 2.0} \\ 
    \toprule
    Methods
    & LSE-C $\uparrow$ 
    & LSE-D $\downarrow$   
    & SIM-O  $\uparrow$ 
    & WER $\downarrow$
    & UTMOS $\uparrow$
    & LSE-C  $\uparrow$ 
    & LSE-D  $\downarrow$  
    & SIM-O  $\uparrow$ 
    & WER $\downarrow$
    & UTMOS $\uparrow$
    \\ 
    \midrule
    GT  & 8.12 &  6.59 & 0.927  &  3.85 & 4.18 & 8.12  & 6.59 & 0.927  & 3.85  &  4.18 \\
    \midrule
    StyleDubber~\cite{cong2024styledubber} (ACL 2024) & 3.87 &  10.92 & 0.607  & 13.14  & 3.14 & 3.74 & 11.00 & 0.501  & 14.18  & 3.04 \\
    Speaker2Dubber~\cite{zhang2024speaker} (MM 2024) & 3.76 & 10.56  & 0.663 & 16.98 &  3.61 & 3.45 &  11.17 & 0.583 & 18.10  & 3.64  \\
    Produbber~\cite{zhang2025produbber} (CVPR 2025) & 2.58 & 12.54  & 0.387 & \textbf{9.45} & 3.85 & 2.78 & 12.14 & 0.310 &  \textbf{11.69} &  3.76 \\
    \midrule
     Ours ($\alpha$ = 0.0) & \textbf{8.21}  &  \textbf{6.89}    &  \textbf{0.754}   &  {9.96}  &  \textbf{3.91}  &  \textbf{8.17}   & \textbf{6.96} & \textbf{0.648}  & {12.95} & \textbf{3.89} \\
    \bottomrule
    \end{tabular}
    }
  \label{result_Chem}%
\end{table*}%

\begin{table}[!tbp]
  \centering
  \caption{The zero shot results under dubbing setting 3.0, which use unseen speaker as refernce audio. } 
  \vspace{-6pt}
  \resizebox{1.0\linewidth}{!}
  {
    \begin{tabular}{lcccccc}
    \toprule
    Methods & LSE-C $\uparrow$ & LSE-D $\downarrow$ & WER $\downarrow$ & UTMOS $\uparrow$ \\
    \midrule
    StyleDubber~\cite{cong2024styledubber} & 6.17  &  9.11 &  15.10 &  3.50 \\
    Speaker2Dubber~\cite{zhang2024speaker} & 4.83 & 10.39
     & 15.91& 3.53 \\
    ProDubber~\cite{zhang2025produbber} & 5.49 & 9.49
     & 14.25 &  3.94 \\
     \midrule 
    Ours ($\alpha$ = 0.0)  & \textbf{7.43} & \textbf{6.64} & \textbf{13.96} & \textbf{3.98} \\
    \bottomrule
    \end{tabular}%
    } 
  \vspace{-6pt}
  \label{tab_setting3_Zero_shot}%
\end{table}

\noindent\textbf{Style Flow Matching Prediction}. Flow matching generates mel-spectrograms $\hat{M}$ from Gaussian noise by a vector field. Given mel-spectrogram space with data $M$, where $M\sim q(M)$. We aim to train a flow matching network to fit $q(M)$ by predicting the probability density path given the vector field, which can be defined as $p_t(x)$. Here $t\in [0,1]$, $p_0(x)= \mathcal{N}(x;0,I)$ and $p_1(x)=q(x)$. Flow matching can predict the probability density path, gradually transforming $x_0\sim p_0(x)$ into $M\sim q(M)$. Our flow matching prediction network is based on optimal-transport conditional flow matching (OT-CFM). OT-CFM uses a linear interpolation flow $\phi_t(x)=(1-(1-\sigma_{\min})t)x_0+tM$, which satisfies the marginal condition $\phi_0(x)=x_0$ and $\phi_1(x)=M$. The gradient field vector field of OT-CFM is $u_t(\phi_t(x)|M)=M-(1-\sigma_{min})x_0$. The training objective of flow matching prediction network is to predict the gradient vector field $v_t(\phi_t(x)|\mu_{SATL},\theta)$, which should be close to $u_t(\phi_t(x)|M)$:
Here $\mu_{SATL}$ is style-enhanced mel-spectrogram level prior according to $\mu$ in Eq.~\ref{eq:mu}. 
To enhance speakers' style, we introduced SATL in flow matching. Specifically, during the flow matching generation process, SATL introduces and enhances style information through an affine transformation, which can be formulated as: 
\begin{equation}
    \mu_{SATL}= \gamma_2(\gamma_1\mu+\beta_1)+\beta_2,
\end{equation}
where $\gamma_1,\gamma_2,\beta_1$, and $\beta_2$ are parameters predicted by SATL based on style features. We train the Style Flow Matching Prediction Network using the condition $\mu_{SATL}$. 
We aim for the Flow Matching prediction network to generate the target mel-spectrogram M conditioned on a given $\mu_{SATL}$. 
During the inference process, the prediction network solves the ODE $d\phi_t(x)=v_t(\phi_t(x)|\mu_{SATL},\theta)dt$ from $t=0$ to $t=1$ to generate a mel-spectrogram $\hat{M}$.

\noindent\textbf{LLM-based Acoustics Flow Matching Guidance}. 
To enhance the clarity of the generated result, we enhanced the mel-spectrograms level prior conditions by LLM-based Acoustics Flow Matching Guidance. 
We observed that the generation process in LLM includes semantic tokens and text tokens, which introduce semantic knowledge. 
Specifically, we enhance LLM's information in flow matching process to improve speech clarity based on classifier-free guidance (CFG), which can be formulated as:
\begin{equation}
    \begin{aligned}
\tilde{v}_t(\phi_t(x)|\mu,\theta)&=v_t(\phi_t(x)|\mu_{SATL},\theta) \\ &+ 
\alpha \Big( v_t(\phi_t(x)|\mu_{SATL},\theta)- v_t(\phi_t(x)|\mu',\theta)\Big),
    \end{aligned}
\end{equation}
where $\mu'=\mathcal{F}([C_{lip}, \mathrm{Up}(\phi, z_p, tab))$, 
and $\phi$ refers to zero vector. 
For adapting dubbing scenarios, our flow matching explicitly decouples the condition inputs into two distinct streams: LLM-based semantic features and original features (aligning with lip movement) to improve dubbing clarity without disturbing the lip aligning prior. 
As a result, we can enhance only the LLM information with classifier-free guidance by controlling the scale factor $\alpha$. 
In general, the proposed guidance mechanism integrates LLM features as high-level semantic conditions to flow-matching network, thereby refining the gradient vector field generation process to ensure clarity while preserving the temporal correlation for audio-visual alignment.

\section{Experimental Results}

\subsection{Implementation Details}
Following the Spark-TTS~\cite{wang2025spark}, the semantic tokenizer consists of 12 ConvNeXt
blocks and 2 downsampling blocks. 
The codebook size of VQ is 8192. 
The ECAPA-TDNN in the global tokenizer features an embedding dimension of 512. 
The cross-modal transformer consists of 8 layers with 2 heads, and the dimension size is 256. 
In dual contrastive aligning, we use 4 heads for multi-head attention with 256 hidden sizes to obtain the attention similarity matrix. 
The temperature coefficient $\tau$ of $\mathcal{L}_{pm}$ and $\mathcal{L}_{mp}$ as both 0.1. 
In data processing, the video frames are sampled at 25 FPS, and all audios are resampled to 16kHz. 
The lip region is resized to 96 $\times$ 96 and pre-trained on ResNet-18, following~\cite{martinez2020lipreading, ma2020towards}. 
The window length, frame size, and hop length in STFT are
640, 1,024, and 160, respectively.  
For LLM-based Acoustics Flow Matching Guidance, the guidance scale is set between 0.0 and 0.8 empirically. 
We set the batch size to 16 on Chem dataset and 64 on GRID.  
Both training and inference are implemented with PyTorch on a GeForce RTX 4090.

\subsection{Datasets}

\noindent\textbf{Chem} is a real-person dubbing dataset recording a chemistry teacher speaking in the class~\cite{prajwal2020learning}. 
It is collected from YouTube, with a total video length of approximately nine hours. 
For complete dubbing, each video has clip to sentence-level~\cite{hu2021neural}. %

\noindent\textbf{GRID} is another real-person dubbing dataset~\cite{cooke2006audio}.  
The whole dataset has 33 speakers, each with 1,000 short English samples. 
All participants are recorded in studio with unified background.

\subsection{Evaluation Metrics}
We abandon some old evaluation metrics and follow the latest speech synthesis technology to evaluate the synthesis quality. 
Specifically, we use LSE-C/D instead of MCD-DTW-SL to evaluate lip-sync. 
We use SIM-O instead of SECS to evaluate speaker similarity.
 We adopt UTMOS instead of MCD-DTW to evaluate quality of speech. 
 Below are the details of each metric:

\noindent\textbf{LSE-C and LSE-D}. 
Compared to the length metric MCD-DTW-SL~\cite{chen2022v2c}, we believe that Lip Sync Error Distance (LSE-D) and Lip Sync Error Confidence (LSE-C)~\cite{chung2016out} can more accurately measure the synchronization of vision and audio. 
These metrics are based on the pre-trained SyncNet~\cite{chung2016out}, which is widely used for lip reading~\cite{YochaiLipVoicer}, talking face~\cite{JiadongSeeing, YoungjoonFaces}, and the video dubbing task~\cite{hu2021neural,lu2022visualtts}. 

\noindent\textbf{SIM-O}. 
To evaluate the timbre consistency between the generated dubbing and the reference audio, we employ the SIM-O following~\cite{N3} to compute the similarity of speaker identity.

\noindent\textbf{UTMOS}.   
UTMOS~\cite{utmos} focuses on evaluating the acoustic quality of synthesized speech~\cite{zhang2025produbber, N3, wang2024maskgct, du2024cosyvoice2, llasa, wang2025spark}, particularly by assessing naturalness, intelligibility, prosody, and expressiveness.

\noindent\textbf{DNSMOS}. 
Deep Noise Suppression MOS (DNSMOS)~\cite{reddy2021dnsmos} is designed to assess the quality of speech processed by noise suppression algorithms, measuring clarity.

\noindent\textbf{SNR score}. 
The signal-to-noise ratio (SNR) score is a deep learning-based estimation system~\cite{snr-score} to assess the clarity of speech. 
A larger SNR corresponds to higher speech clarity. 

\noindent\textbf{WER}. 
The Word Error Rate (WER)~\cite{AndrewWER} is used to measure pronunciation accuracy by using Whisper-V3~\cite{whisper} as the ASR model.

\begin{table*}[!t]
  \centering
    \caption{
    Compared with related Dubbing methods on GRID benchmark 
    under the same dub setting as the Chem benchmark. 
    }
    \vspace{-6pt}
  \resizebox{1.0\linewidth}{!}
  {
    \begin{tabular}{c|ccccc|ccccc}
    \hline
    Setting & \multicolumn{5}{c|}{Dubbing Setting 1.0} & \multicolumn{5}{c}{Dubbing Setting 2.0} \\ 
    \toprule
    Methods
    & LSE-C $\uparrow$ 
    & LSE-D $\downarrow$   
    & SIM-O $\uparrow$ 
    & WER $\downarrow$
    & UTMOS $\uparrow$
    & LSE-C  $\uparrow$ 
    & LSE-D  $\downarrow$  
    & SIM-O $\uparrow$ 
    & WER $\downarrow$
    & UTMOS  $\uparrow$
    \\   
    \midrule
    GT & 7.13 & 6.78 & 0.866 & 0.00  &  3.94 & 7.13 & 6.78 & 0.866 & 0.00  &  3.94  \\
    \midrule
    StyleDubber~\cite{cong2024styledubber} (ACL 2024) & 6.12 & 9.03 & 0.754 & 18.88  & 3.73  & 6.09 & 9.08 & 0.617 & 19.58 & 3.71  \\
    Speaker2Dubber~\cite{zhang2024speaker} (MM 2024) & 5.27 & 9.84 & 0.734 & \textbf{17.04} & 3.69  & 5.19 & 9.93 & 0.606 & \textbf{17.00}  &  3.73 \\
     Produbber~\cite{zhang2025produbber} (CVPR 2025) & 5.23 &  9.59 &  0.791 & 18.60 &  3.87 & 5.56 & 9.37 & 0.663  & 19.17  & 3.86 \\
    \midrule
     Ours ($\alpha$ = 0.0) & \textbf{7.27}  &  \textbf{6.72}     &  \textbf{0.811}   &  18.54  &  \textbf{3.97}  &  \textbf{7.20}   & \textbf{6.75}  &\textbf{0.679}  & {19.24} &  \textbf{3.95} \\
    \bottomrule
    \end{tabular}
    }
  \label{result_Grid}
\end{table*}%

\begin{table}[!tbp]
  \centering
  \caption{The Clarity performance of using different scale $\alpha$ in acoustics flow matching guidance. 
  Note that DNSMOS, SNR Score, and UTMOS are not human subjective metrics. } 
  \vspace{-6pt}
    \begin{tabular}{lccc}
    \toprule
    Guidance Scale & DNSMOS $\uparrow$ & SNR Score $\uparrow$  & UTMOS $\uparrow$  \\
    \midrule
    Produbber~\cite{zhang2025produbber} & 3.664 &  23.703 &  3.849  \\
    \midrule
    Ours ($\alpha$ = 0.0)   & 3.745 & 26.341  & 3.912   \\
    Ours ($\alpha$ = 0.2)  & 3.777 & 26.657 & 3.929     \\
    Ours ($\alpha$ = 0.4)  & 3.799 & 26.706 & 3.940    \\
    Ours ($\alpha$ = 0.6)  & 3.819 & 26.903 & 3.953    \\
    Ours ($\alpha$ = 0.8)  & \textbf{3.829} & \textbf{27.016}  & \textbf{3.960}    \\
    \bottomrule
    \end{tabular}%
    \vspace{-6pt}
  \label{Voice_Enhancement}%
\end{table}

\subsection{Comparison with SOTA Dubbing Methods}

\noindent\textbf{Results on the Chem Dataset}. 
As shown in Table~\ref{result_Chem},
our method achieves the best performance on almost all metrics on the Chem benchmark, whether in setting 1 or setting 2.  
First, our method achieves
the best LSE-C and LSE-D, with absolute improvements of
5.63\% and 5.65\% than the TTS-based dubbing methods with duration predictor (like StyleDubber~\cite{cong2024styledubber}, Speaker2Dubber~\cite{zhang2024speaker}, Produbber~\cite{zhang2025produbber}),  demonstrating the effectiveness of our methods in lip-sync by LLM-based semantic-aware learning and dual contrastive aligning.  
Besides, the dubbing synthesis quality of our method is the highest among all dubbing methods, with a UTMOS score of 3.91. 
In summary, \dubber~ is a comprehensive dubbing model that makes up for the shortcomings of previous methods in audio-visual synchronization, speaker similarity, and dubbing synthesis quality, and achieves a WER comparable to SOTA.

\noindent\textbf{Results on the GRID Dataset}. 
As shown in Table~\ref{result_Grid}, a similar trend is found in the multi-speaker benchmark. 
We still achieve SOTA performance in audio-visual synchronization, dubbing synthesis quality, and discrepancy from ground truth in both dubbing settings while maintaining similarity with advanced speaker identity. 
Specifically, our method can achieve similar WER as ProDubber~\cite{zhang2025produbber} while maintaining higher LSE-C/D than previous TTS-based dubbing methods (like StyleDubber, Speaker2Dubber, Produbber), which adopt a Duration Predictor (DP) to produce rough duration, leading to poor audio-visual alignment. 
Finally, the UTMOS of our method is improved by 12\% over Speaker2Dubber~\cite{zhang2024speaker} on setting 2, which shows that the speech quality synthesized by our method is the best, even better than the two-stage pre-training manner.

\noindent\textbf{Results on the Speaker Zero-shot Test}. 
In addition to dubbing benchmarks, we also conduct the zero-shot test to evaluate the generalization performance of models. 
This setting uses the audio of unseen characters (from another dataset) as reference audio. 
Here, we use the audio from the Chem dataset as reference audio to measure the GRID dataset. 
As shown in Table~\ref{tab_setting3_Zero_shot}, our proposed method surpasses the current state-of-the-art models and achieves the best performance across all metrics. 
Besides, we still achieve the best lip-sync (see LSE-C and LSE-D) in zero-shot setting.

\begin{table}[!tbp]
  \centering
  \caption{Ablation study of the proposed method on the Chem benchmark dataset with 1.0 setting.} 
  \vspace{-6pt}
  \resizebox{1.0\linewidth}{!}
  {
    \begin{tabular}{lccccccc}
    \toprule
    \# & Methods   & LSE-C $\uparrow$ & LSE-D $\downarrow$ & WER $\downarrow$ & SIM-O $\uparrow$ & UTMOS $\downarrow$  \\
    \midrule
    1 & w/o FVE  & 8.18  & 6.94 & 13.85  & 0.620 &  3.66 \\
    2 & w/o LLM-SL & 8.16 & 6.95 & 48.33  & 0.671 & 3.76  \\ 
    3 & w/o DCA  & 3.62  & 10.28 &  10.04 & 0.747 & {3.90} \\
    4 & w/o Style in FVE  & 8.19  & 6.92 & 14.96  & 0.582 & 3.84  \\
    \midrule
    5 & Full model  & \textbf{8.21}  &  \textbf{6.89}  &  \textbf{9.96} & \textbf{0.754}  & \textbf{3.91}  \\ 
    \bottomrule
    \end{tabular}%
    } 
  \label{tab_ablation}%
\end{table}

\subsection{Analysis of Flow-based Voice Enhancing} 
As shown in~\ref{Voice_Enhancement}, we use DNSMOS, SNR, and UTMOS as main metrics. 
As the guidance scale increases, DNSMOS, SNR, and UTMOS all show improvement, indicating that LLM-based Acoustics Flow Matching Guidance effectively reduces noise and enhances speech clarity and overall intelligibility.  
Besides, we find that DNSMOS increases faster than UTMOS, indicating that the proposed method primarily enhances clarity.

\begin{figure*}[!tbp]
\begin{center}
\includegraphics[width=1\textwidth]{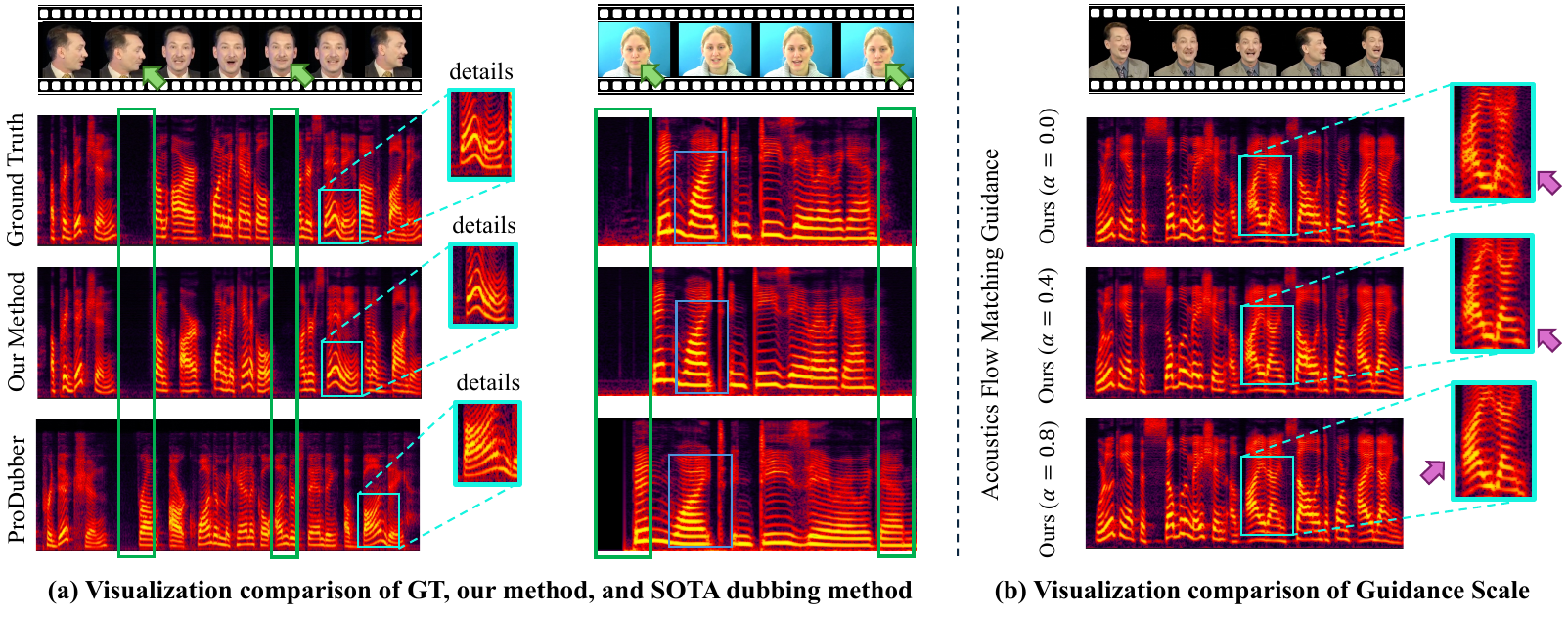}
\end{center}
\vspace{-0.2cm}
   \caption{The visualization of the mel-spectrograms of ground truth (GT) and synthesized audios obtained by different models. %
   In~(a), green arrows point to the video frames that do not speak, and green bounding boxes are used to highlight the pauses in speech. 
   In~(b), pink arrows point to the enhanced details of the mel-spectrogram as flow matching guidance scale $\alpha$ increases. 
   }
\label{Visualall}
\end{figure*}

\subsection{Ablation Studies}
To further investigate the specific effects of main module in our method, we conduct ablation studies on the Dub 1.0 setting of the Chem benchmark. 
The ablation results are presented in Table~\ref{tab_ablation}. 
It shows that all modules contribute significantly to the overall performance, and each module has a different focus.  
When LLM-SL is removed, both WER and UTMOS decrease, with WER being more obvious. 
This shows that LLM-based semantic-aware learning can provide rich semantic information on phoneme level, which is necessary for clear pronunciation. 
When removing DCA and using the duration predictor to provide alignment, we observe a significant degradation in LSE-C and LSE-D. 
Last, removing Style in FVE has a greater impact on speaker similarity (see SIM-O).

\subsection{Compare with Different Audio Generators} 

Please note that when comparing with the dubbing baseline (Table~\ref{result_Chem}-\ref{tab_ablation}), we adopt HiFi-GAN~\cite{Kong2020HiFi} as audio generator to convert the mel-spectrogram to waveforms to ensure fairness. 
To explore the upper-bound quality of the generated audio by using different audio generators, we select more powerful audio generators: BigVGAN~\cite{SangBigVGAN}, 16K Hz Descript Audio Codec (DAC)~\cite{DAC}, and 24K Hz Codec Vocoder (CV)~\cite{du2024cosyvoice2}, respectively. 
To ensure the integrity of the original design (without removing the FVE module), we do not consider directly decoding waveforms from tokens. Therefore, for DAC and CV, we first generate the original waveform and then perform reconstruction. 
As shown in Table~\ref{Compare_with_AudioDeocder}, the results show that 24K CV achieves the best speech quality (see UTMOS), while BigVGAN achieves better alignment and timbre restoration with a slight advantage. 
Most importantly, we find that all audio generators are better than SOTA dubbing baseline (\eg, Produbber~\cite{zhang2025produbber}) or powerful TTS methods (see Table~\ref{Compare_with_TTS}) in audio-visual synchronization (see LSE C/D), because the aligning information has been preserved in advance.  
This is also the advantage of our design, which can be extended by stronger audio generators in the future.

\begin{table}[!tbp]
  \centering
  \caption{Compared with different audio generators. The results under the $\alpha$=0.8  guidance scale of FVE.} 
  \vspace{-6pt}
  \resizebox{1.0\linewidth}{!}
  {
    \begin{tabular}{lccccccc}
    \toprule
    Methods & Type & LSE-C $\uparrow$ & LSE-D $\downarrow$ & SIM-O  $\uparrow$ &  UTMOS $\uparrow$ \\
    \midrule
    Ours (HiFiGAN) & mel. & {8.163} & {6.954} & {0.745} & 3.960 \\
    Ours (BigVGAN) & mel. & \textbf{8.185} & \textbf{6.932} & \textbf{0.749} & 3.971 \\
    Ours (16K DAC) & codec  & {8.101} & {6.980} & {0.703} & 3.916  \\
    Ours (24K CV) & codec & {8.179} & {6.958} & {0.721} & \textbf{4.154}  \\
    \bottomrule
    \end{tabular}%
    } 
    \vspace{-6pt}
  \label{Compare_with_AudioDeocder}%
\end{table}

\subsection{Compare with LLM-based TTS method}

As shown in Table~\ref{Compare_with_TTS}, we compare with the recent LLM-based TTS methods. 
Our method achieves the best performance in LSE-C and LSE-D to maintain synchronization, which is extremely important for moving towards automated lip-sync dubbing. 
Besides, our dubbing scheme can approach or even exceed part of large-scale TTS methods in UTMOS. 
For example, our UTMOS is 3.59\% higher than FireRedTTS. 
In contrast, most part of LLM-based TTS methods cannot adapt to dubbing scenes due to the lower LSE-D and LSE-C, proving the bad audio-visual alignment with lip movement.

\subsection{Qualitative Analysis} 

We visualize the mel-spectrograms of ground truth and dubbing generated by different models for comparison in Figure~\ref{Visualall}.  
The green bounding boxes highlight the pauses in the speech, and blue bounding boxes exhibit significant differences in acoustic details. 
We have also enlarged the details to make it easier for readers to compare. 
As shown in Figure~\ref{Visualall}(a), 
our method demonstrates high-quality audio-visual alignment and acoustic quality relative to state-of-the-art dubbing baseline. 
In the corresponding silent video frames (see green arrows), our method can generate the same sound pauses as GT, which illustrates the effectiveness of dual contrastive aligning. 
As shown in Figure~\ref{Visualall}(b), we visualize the mel-spectrogram generation effect of Acoustics Flow Matching Guidance at different scales. 
As the scale increases, the originally blurry and artifact-filled spectrum gradually becomes clearer. 
The qualitative analysis shows that our model can generate high-quality audio-visual alignment and high-fidelity acoustic quality.

\begin{table}[!tbp]
  \centering
  \caption{Compared with SOTA LLM-based TTS method.} 
  \vspace{-6pt}
  \resizebox{1.0\linewidth}{!}
  {
    \begin{tabular}{lcccccc}
    \toprule
    Methods & Dub. & LSE-C $\uparrow$ & LSE-D $\downarrow$ & SIM-O $\uparrow$ & UTMOS  $\uparrow$ \\
    \midrule
    CosyVocie 2.0~\cite{du2024cosyvoice2} & $\times$ & 3.001 & 12.248 & 0.718  & 4.252 \\
    Llasa-3B~\cite{llasa} & $\times$ & 3.537 & 11.564 & 0.662 & 4.207 \\
    Spark-TTS~\cite{wang2025spark}  & $\times$ & 2.850 & 12.347 & 0.549 & \textbf{4.390} \\
    FireRedTTS~\cite{guo2024fireredtts} & $\times$ & 2.779  & 12.413 & 0.529 & 4.010 \\
     \midrule 
    Ours (24K CV) & $\checkmark$  & \textbf{8.179} & \textbf{6.958} & \textbf{0.721} & 4.154 \\
    \bottomrule
    \end{tabular}%
    } 
    \vspace{-12pt}
  \label{Compare_with_TTS}%
\end{table}

\section{Conclusion}
In this paper, we propose an LLM-based dubbing architecture, which incorporates a large language model for semantic-aware learning and voice-enhanced flow matching for acoustic modeling. 
By LLM-based semantic-aware learning, the model absorbs the phoneme-level semantic knowledge with in-contextual information, while maintaining the lip-sync by dual contrastive aligning. 
Besides, the flow-based voice enhancing ensures the acoustic clarity and speaker identity. 
Our proposed model sets SOTA results on both Chem and GRID benchmarks. 
In the future, we will explore the wild datasets and provide lightweight solutions to perform fast inference.

\section*{Acknowledgments}


This work was supported by the National Nature Science Foundation of China (62322211), the "Pioneer" and "Leading Goose" R\&D Program of Zhejiang Province (2024C01023), Key Laboratory of Intelligent Processing Technology for Digital Music (Zhejiang Conservatory of Music), Ministry of Culture and Tourism (2023DMKLB004). 
Amin Beheshti, Anton van den Hengel, and Yuankai Qi are not supported by the aforementioned fundings. 


\bibliographystyle{ACM-Reference-Format}
\balance
\bibliography{sample-base}

\end{document}